\documentclass[12pt,oneside,a4paper]{article}
\usepackage{fancyhdr, ifpdf}
\usepackage{syntonly}
\usepackage{geometry}
\usepackage{makeidx}
\makeindex
\usepackage[english]{babel}
\newtheorem{definition}{Definition}
\newtheorem{theorem}{Theorem}

\newtheorem{lemma}{Lemma}
\newtheorem{corollary}{Corollary}
\newtheorem{proposition}{Proposition}

\usepackage{makeidx}
\makeindex

\usepackage{amsmath}
\usepackage{amssymb}
\usepackage{amsmath}
\usepackage{amsfonts}
\usepackage{amssymb}
\usepackage{diagxy}

\title{Algorithmic randomness and Ramsey properties of countable homogeneous structures.}

\author{Willem L. Fouch\'{e} \footnote{The research is based upon work supported by the National Research Foundation (NRF) of South Africa. Any opinion, findings and conclusions or recommendations expressed in this material are those of the author and therefore
the NRF does not accept any liability in regard thereto.} \footnote{email:fouchwl@gmail.com}\\
\it Department of Decision Sciences,\\ \it School of Economical Sciences,\\\it University of South
Africa,  Pretoria, South
Africa.}
\date{}

\begin{document}
 \maketitle

\pagenumbering{arabic}
\abstract\small{We  study, in the context of algorithmic randomness, the closed amenable subgroups of the symmetric group $S_\infty$ of a countable set.  In this paper we address this problem by investigating  a  link between the symmetries associated with  Ramsey   Fra\"iss\'e order classes  and algorithmic randomness.}

Fra\"iss\'e limits, Ramsey theory.

\newcommand{\qed}{\nobreak \ifvmode \relax \else
      \ifdim\lastskip<1.5em \hskip-\lastskip
      \hskip1.5em plus0em minus0.5em \fi \nobreak
      \vrule height0.30em width0.4em depth0.25em\fi}

\section{Introduction}

The focus of this work is , as for example in \cite{FouchPot, Fouchrandomizer}, on the  problem of understanding the symmetries that transform a recursively presented  universal structure, which in this paper will be a Fra\" {i}ss\' e limit of finite first order structures, to a copy of such a structure which is Martin-L\" of random relative to a canonical  $S_\infty$-invariant measure on the class of all universal structures of the given type. Here $S_\infty$ is  the symmetric group of a countable set, with the pointwise convergence topology.
This invesigation leads to a  link between the symmetries associated with the so-called discernable flows in structural Ramsey theory and algorithmic randomness.

Glasner and Weiss \cite{GlasnerWeiss} showed that there exists a unique measure on the set of linear orderings of the natural numbers (seen as a subset of Cantor space) that is invariant under the canonical action of the symmetric group of the natural numbers.
The author \cite{Fouchrandomizer} showed that this measure is computable and studied the associated Martin-L\" of (ML) random points, which, due to the uniqueness of the Glasner-Weiss measure, may be regarded as random linear orders. 
The author showed that any ML-random linear order has the order type of the rationals. Moreover, it was shown that recent work by Kechris and Sokic \cite{KechrisSokic} implies that no random linear order can be the extension of the universal poset  (the Fra\"iss\'e limit of finite posets) to a linear order. In \cite{Fouchrandomizer} a study was made of  so-called ``randomizers". These are permutations of the natural numbers that transform a computable (Cantor) rational linear order into a random one. It was also  proven in \cite{Fouchrandomizer} that any such randomizer cannot be an automorphism of the universal poset.

The aim of this paper is to generalise these results to a broader class of Fra\" {i}ss\' e limits  $\mathbb{F}_0$ of  Ramsey classes, the automorphism groups $\mbox{Aut}(\mathbb{F}_0)$  not being amenable. Again, as in \cite{Fouchrandomizer},  this paper relies heavily on the groundbraking paper \cite{Kechrisetal} by Kechris, Pestov and Todorcevic. The arguments in this paper  require some  understanding of  the subtle interplay between structural Ramsey theory and topological dynamics as is beautifully explicated in the paper \cite{Kechrisetal}. This paper has been written  in such a way that it should be accessible to a non-specialist in Ramsey theory.

\section{Preliminaries on amenable groups}
\label{sec:amenable groups}

Let $G$ be a topological group and $X$ a compact Hausdorff space. A dynamical
system $(X,G)$ (or a $G$-flow on $X$) is given by a jointly continuous action of $G$ on $X$. If $(Y, G)$ is
a second dynamical system, then a $G$-morphism  $\pi : (X, G) \rightarrow  (Y, G) $
is a continuous mapping $\pi:X \rightarrow Y$ which intertwines the  $G$-actions, i.e.,the diagram

\[\bfig
\square[G \times X`X`G\times Y`Y; \alpha`1\times\pi`\pi`\beta  ]
   \efig ,\]    
commutes with $\alpha,\beta$ being the group actions.

 An isomorphism is a bijective homomorphism. A subflow of $(X,G)$ is a $G$-flow on a compact subspace $Y$ of $X$ with  the action of $G$ on $X$ restricted to the action on $Y$. A  $G$-flow is minimal if it has no proper subflows. Every dynamical system has a minimal subflow (Zorn).                         

The following fact, first proven by Ellis (1949) \cite{Ellis}, is central to the theory of dynamical systems:
\begin{theorem}
Let $G$ be a Hausdorff topological group. There exists, up to \\$G$-isomorphism, a unique minimal dynamical system, denoted by $(M (G),G)$, such that   for every minimal dynamical system $(X, G)$ there exists a $G$--epimorphism $$\pi : (M, G) \longrightarrow
(X, G),$$ and any two such universal systems are isomorphic.
\end{theorem}
The flow $(M (G),G)$ is called the {\it universal minimal flow} of $G$.

 We next introduce the notion of amenable groups.
\begin{definition}
A topological group $G$ is {\it amenable} if, whenever $X$ is a non-empty compact Hausdorff space and $\pi$ is a continuous action of $G$ on $X$, then there is a $G$--invariant Borel probability measure on $X$.
\end{definition}
This means that, for every $G$-flow on a compact space $X$,  there is a measure $\nu$ on the Borel algebra of $X$, such that, $ \nu(X)=1$ and, for every $g \in G$ and Borel subset $U$ of $X$,
$$\nu(gU)=\nu(U).$$

\section{Fra\" {i}ss\' e limits and their recursive representations}
\label{sec:fraisse limits}
In the sequel, ${\cal L}$ will stand for the 
signature of 
a relational 
structure. Moreover, ${\cal L}$ will always be finite and 
the arities 
of the relational 
symbols will all be $\geq 1$.
 The definitions that follow were 
introduced by 
Fra\"\i ss\'e 
in 1954. 

The {\em age} of an ${\cal L}$-structure $X$, written $\mbox{\tt Age}( X )$, is 
the class of all finite 
${\cal L}$-structures (defined on finite ordinals) 
which can 
be embedded as ${\cal 
L}$-structures into $X$. The structure $X$ is {\em 
homogeneous} (
some authors say {\em
ultrahomogeneous} ) if, given any isomorphism $f: A \rightarrow
B$ between 
finite substructures of 
$X$, there is an automorphism $g$ of $X$ whose restriction 
to $A$ is 
$f$. 
A class {\bf K} of finite ${\cal 
L}$-structures has the 
{\em amalgamation 
property} if, for structures $A, B_1, B_2$ in {\bf K} and 
embeddings 
$f_i : A \rightarrow B_i$ 
($i=1,2$) there is a structure $C$ in {\bf K} and there are 
embeddings 
$g_i : B_i \rightarrow C$ 
($i=1,2$), such that the following diagram  commutes: 

   \[\bfig
   \morphism(0,0)|l|<250,-250>[A` B_2;f_2] 
   \morphism(0,0)<250,250>[A`B_1;f_1] 
   \morphism(250,250)/{-->}/<250,-250>[B_1`C;g_1]
    \morphism(250,-250)|r|/{-->}/<250,250>[B_2`C ;g_2] 
      \efig.\]
Suppose {\bf K} is a countable class of finite ${\cal 
L}$-structures, 
the domains of which are 
finite ordinals such that 
\begin{enumerate}
\item if $A$ is a finite ${\cal L}$-structure defined on 
some finite 
ordinal,  if $B \in 
\mbox{\bf K}$ and if there is an embedding of $A$ into $B$, 
then $A \in 
\mbox{\bf K}$;
\item the class {\bf K} has the amalgamation property.
\end{enumerate}
Then, Fra\"\i ss\'e showed that there is a countable 
homogeneous 
structure $X$ such that 
$\mbox{\tt Age}(X) = \mbox{\bf K}$. Moreover, $X$ is unique up to 
isomorphism. 
The (essentially) unique $X$ is called 
the {\em Fra\"\i ss\'e limit}  $\mathbb{K}$ of {\bf K}. Note
that, 
conversely, the age {\bf K} of 
a countable homogeneous structure has properties (1) and (2). We shall frequently call a countable structure which is isomorphic to a Fra\" {i}ss\' e limit a {\em universal structure}.

 A {\em recursive representation} of a countably infinite 
${\cal 
L}$-structure 
$X$ is a bijection $\phi: $X$ \rightarrow \mathbb{N}$ such that, for 
each $R \in 
{\cal L}$,
if the arity of $R$ is $n$, then the relation $R^{\phi}$ 
defined on 
$\mathbb{N}^n$ by
$$R^{\phi} \left( x_1, x_2, \ldots, x_n \right) 
\Longleftrightarrow R 
\left( \phi^{-1}(x_1), 
\ldots, \phi^{-1}(x_n) \right),$$
is recursive. If we identify the underlying set of $X$ with 
$\mathbb{N}$ 
via $\phi$ and each $R$ 
with $R^{\phi}$, we call the resulting structure a {\em 
recursive ${\cal 
L}$-structure} on $\mathbb{N}$ and we say it has a recursive representation on $\mathbb{N}$.

If $X$ is countable and homogeneous and if $\mbox{\tt Age} (X)$ has an 
enumeration $A_0$, $A_1$, $A_2$, 
$\ldots $,
possibly with repetition, with the property that there is a 
recursive 
procedure that 
yields, for each $i \in \mathbb{N}$, and $R \in {\cal L}$, the 
underlying set 
$A(i)$ of 
$A_i$ together with the interpretation of $R$ in $A(i)$, 
then we call 
$\left( A_i : i  \in \mathbb{N}  \right)$ a recursive enumeration of 
$\mbox{\tt Age} (X)$ . It 
follows from the construction of Fra\"\i ss\'e limits from 
their ages, 
that one can construct a recursive 
representation of $X$ 
from a recursive enumeration of its age. (Conversely, 
it is 
trivial to 
derive a recursive enumeration of $\mbox{\tt Age} (X)$ from a recursive 
representation of $X$.) 
It is therefore not difficult to find recursive 
representations for 
Fra\"\i ss\'e 
limits of classes {\bf K} from recursive enumerations of 
their ages.
\begin{theorem}
Suppose $\mathbb{C}$ and $\mathbb{D}$ are countable recursively represented   $\cal{L}$-structures on $\mathbb{N}$ with the same age. Suppose that they are both homogeneous. Then there is a recursive isomorphism from $\mathbb{C}$ to $\mathbb{D}$.
\label{th:efffraisse}
\end{theorem}
Proof.  As mentioned in \cite{Fouchrandomizer}, the model-theoretic back-and-forth argument as  discussed, for example, on pp 161-162 of Hodges \cite{Hodges} is constructive relative to the recursive representations of the homogeneous structures $\mathbb{C}$ and $\mathbb{D}$. 
\section{Structural Ramsey theory in a model theoretic context}

In this section we summarise the results from \cite{Kechrisetal} which underly the formalisation and proof of the main theorem of this paper. Unless otherwise stated all the proofs of the statements made here can be found in \cite{Kechrisetal}.

Let $\mathbf{K}$ be the age of some countable $\cal{L}$-structure. For $A, \pi \in \mathbf{K}$ we denote by $A^\pi$ the set of all the (model-theoretic)  structure-preserving  embeddings of $\pi$ in $A$. For a natural number $r\geq 1$ and for $\pi, A, B \in \mathbf{K}$ we introduce the predicate $B \leadsto\left(A\right)_r^\pi$ (Erd\H os-notation) to mean:
\[ B \leadsto\left(A\right)_r^\pi
\Longleftrightarrow
\big(
\forall_{\bfig
                  \morphism(0,0)/{->}/<250,0>[B^\pi`r;\chi]
                 \efig} 
\exists_{\bfig
\morphism(0,0)/{>->}/<250,0>[A`B;\alpha]
\efig}
\bfig
         \morphism(0,0)<500,0>[A^\pi`B^\pi;\alpha_*]
          \morphism (0,0)|l| <250,-250>[A^\pi`r; !]
           \morphism(500,0)|r|/{->}/<-250, -250>[B^\pi `r; \chi]
            \morphism(0,0)/{}/<250,-100>[A^\pi`\oplus;]
          \efig 
\big).\]

Here $\alpha_*:A^\pi \rightarrow B^\pi$ is the mapping that takes an embedding $ \bfig
\morphism(0,0)/{>->}/<250,0>[\pi`A;x]
\efig$ to the induced embedding  $ \bfig
\morphism(0,0)/{>->}/<250,0>[\pi`B;\alpha x]
\efig$.

In other words, $ B \leadsto\left(A\right)_r^\pi$ iff:  for every $r$-colouring $\chi$ of the set $B^\pi$ consisting of the embeddings of $\pi$ in $B$ (copies of $\pi$ in $B$), there is an embedding $\alpha$ of $A$ into $B$   such that $\chi\alpha_*$ is a constant. This means that $\chi$ assumes a constant value on all the embeddings of $\pi$ into the image $A' \subset B$ of $A$ under $\alpha$. 

We shall call an age $\mathbf{K}$ a {\em Ramsey age} if, for all $\pi,A \in \mathbf{K}$ with $A^\pi \neq \emptyset$, and all natural numbers $r\geq 1$, there is some $B \in \mathbf{K}$ such that $B \leadsto\left(A\right)_r^\pi$.

Assume $\mathcal{L}$ is a countable  signature containing a distinguished binary relation
symbol $<$. An {\it order structure} $A$ for the signature $\mathcal{L}$ with the distinguished symbol $<$, is a structure $ A$ for which  the interpretation $<^A$ of  the  symbol $<$  in $A$ is
a total ordering.

 An {\it order class} $\mathbf{K}$ for $\mathcal{L}$ is one for which all $A \in \mathbf{K}$ are
order structures (relative to the distinguished $<$).

Let $\mathcal{L}_0$ be the signature obtained by removing the  distinguished symbol $<$ from $\mathcal{L}$.  For any $\mathcal{L}$-structure $A$, denote by $A_0$  the $\mathcal{L}_0$-structure  which is the reduct of $A$ to $\mathcal{L}_0$. This means that $A_0$  is the structure $A$ where the distinguished order $<$ interpretated as a total order $<^A$  in $A$ is being ignored.

Let $\mathbf{K}$ be a Fra\"\i ss\'e  order class. Denote by $\mathbf{K}_0$ the class of all reducts $A_0$ for some $ A \in \mathbf{K}$. Write $\mathbb{F}$ for the  Fra\"\i ss\'e  limit of $\mathbf{ K}$. We now discuss when $\mathbf{K}_0$ is also a  Fra\"\i ss\'e  class with limit $\mathbb{F}_0$ the latter being the reduct of $\mathbb{F}$ to $\mathcal{L}_0$.
Following Kechris, Pestov and Todorcevic \cite{Kechrisetal}, we say that the class $\mathbf{K}$ is {\it reasonable} if for every $A_0, B_0 \in \mathbf{K}_0$, and linear ordering $<$ on $A_0$ such that $(A_0,<) \in \mathbf{K}$, and for an embedding $\pi: A_0 \rightarrow B_0$,  there is a linear ordering $<_1$   on $B_0$ so that $B=(B_0,<_1,) \in \mathbf{K}$ and $\pi : A \rightarrow B$ is also an embedding.  (This means that $x <y \Leftrightarrow \pi(x) <_1 \pi(y)$.) \\\\Then Kechris et al (p 135) showed that $\mathbf{K}_0$ is a  Fra\"\i ss\'e  class with limit $\mathbb{F}_0$, (which is the reduct of $\mathbb{F}$ to the signature $\mathcal{L}_0$)  iff the  Fra\"\i ss\'e order class $\mathbf{K}$ is reasonable. \\\\Note that, in this case, the underlying sets of $\mathbb{F}$ and $\mathbb{F}_0$ are the same.  Moreover, we can write $$\mathbb{F} =(\mathbb{F}_0, <_0),$$ for some linear ordering $<_0$ on the underlying set of $\mathbb{F}_0$.

 We consider the  continuous action of the automorphism group $\mbox{Aut}(\mathbb{F}_0)$ on the (topological)  space of all linear orderings on the set $F_0$, which is the underlying set of the structure $\mathbb{F}_0$. Write $X_{\mathbf{K}} \subset \{0,1\}^{F_0 \times F_0}$ for the orbit topological closure  of the action of $\mbox{Aut}(\mathbb{F}_0)$ on the linear ordering $<_0$, i.e., 
$$X_{\mathbf{K}} =\overline {\mbox{Aut}(\mathbb{F}_0).<_0}.$$
This set is clearly a closed, hence compact, subset of the Baire space $\{0,1\}^{F_0 \times F_0}$. Moreover, it is clearly also  an $\mbox{Aut}(\mathbb{F}_0)$-invariant subset of $\{0,1\}^{F_0 \times F_0}$ under the natural action of $\mbox{Aut}(\mathbb{F}_0)$ on the latter space.
We have thus obtained an $\mbox{Aut}(\mathbb{F}_0)$-flow on $X_{\mathbf{K}}$ . \newline
This flow can be defined for any reasonable (in the technical sense as explained above)  Fra\"\i ss\'e order class $\mathbf{K}$.  I will call it the {\it discerning } flow associated with the  reasonable Fra\"\i ss\'e order class $\mathbf{K}$.\newline
{\bf Remark.} The author uses the terminology {\it discerning} in acknowledgement of Ramsey´s pioneering work in developing his theorem in the context what now would be considered as a study of indiscernables in model theory.

If, in addition to being a Fra\"\i ss\'e  order class, the class $\mathbf{K}$ is Ramsey, then every minimal subflow of the discernable flow is isomorphic to the universal minimal flow of $\mbox{Aut}(\mathbb{F}_0)$.

The Fra\"\i ss\'e order class $\mathbf{K}$  is said to have the {\it ordering property} if for every $A_0 \in \mathbf{K}_0$, there is a $B_0 \in \mathbf{K}_0$ such for any linear ordering $<$ on $A_0$ and every linear ordering $<_1$ on $B_0$, where both $<,<_1$ are restrictions of $<_0$, there is an embedding of $(A_0,<)$ into $(B_0,<_1)$.

The discerning flow associated with the Fra\"\i ss\'e order class $\mathbf{K}$ is itself minimal iff $\mathbf{K}$ has the ordering property. 

We also extract the following remark from \cite{Kechrisetal}.
\begin{proposition}
If $\mathbf{K}$ is a  Fra\"\i ss\'e order class which is Ramsey and has the ordering property, then a total order $\xi$ belongs to the discerning flow  $X_\mathbf{K}$ iff for any $A$  in the age of $\mathbb{F}_0$ it is the case that $<^A$ is the restriction of $\xi$ to $A$.
\label{lemmacrux}
\end{proposition}
We shall make substantial use of this remark in the sequel.
\newline
{\bf Example.}
It is known (see, for example \cite{Fouchposet} that the class $\mathbf{P}$ (finite posets, linear extensions) is Ramsey and has the ordering property.  As was noted in \cite{Kechrisetal}, this has the implication that the discerning $\mbox{Aut}(\mathbb{P}_0)$-flow is thus a universal minimal flow. It acts on the space $X_{\mathbf{P}}$ consisting of the linear extensions of the universal poset $\mathbb{P}_0$. Using these facts, Kechris and Soki\v c (2011) \cite{KechrisSokic} recently showed that  the automorphism group of $\mathbb{P}_0$ is not amenable. These results do imply that the set of linear extensions of the Fra\" {i}ss\' e  limit  of finite posets  are all, in a definite sense, nonrandom, at least from the point of view of algorithmic randomness as was shown in \cite{Fouchrandomizer}. This result will be placed in a broader context in what follows.

\section{Martin-L\"of random countable orders}
\label{sec:random countable orders}


Let $S_\infty$ be the group of permutations  of a countable set, which, without loss of generality, we may take to be  $\mathbb{N}$. We place on $S_\infty$ the pointwise convergence topolopy.
Let $(\mathbb{N}\times\mathbb{N})_{\neq}$ denote the set of ordered pairs $(i,j)$ of natural numbers with $i \neq j$. Write $\mathcal{M}$ for the set of total orders on  $\mathbb{N}$. We identify $\mathcal{M}$ with a subset of $\{0,1\}^{(\mathbb{N}\times\mathbb{N})_{\neq}}$ by identifying a total order  $<$  on $\mathbb{N}$ with the function $\xi:(\mathbb{N}\times\mathbb{N})_{\neq} \rightarrow \{0,1\}$ given by
$$\xi(x,y)=1 \Leftrightarrow x < y,\;\;x,y \in \mathbb{N}.$$ 
The total order associated with $\xi$ will be denoted by $<_\xi$.
We topologise $\mathcal{M}$ via the natural injection $$\mathcal{M} \longrightarrow \{0,1\}^{(\mathbb{N}\times\mathbb{N})_{\neq}},$$ where the (Baire) space $\{0,1\}^{(\mathbb{N}\times\mathbb{N})_\neq}$ has the product topology. As such $\mathcal{M}$ is a closed hence compact subspace of $\{0,1\}^{(\mathbb{N}\times\mathbb{N})_\neq}$.

The group $S_\infty$ acts continuously on $\mathcal{M}$ if, for $\xi \in \mathcal{M}$ and $\sigma \in S_\infty$, we define the total order $\sigma\xi$ by:
\[x <_{\sigma\xi} y \Longleftrightarrow \sigma^{-1}x <_\xi \sigma^{-1}y,\;x,y \in \mathbb{N}.\]

Since $S_\infty$ is an amenable group, there is an $S_\infty$-invariant measure on $\mathcal{M}$. In fact, Glasner and Weiss (2002) \cite{GlasnerWeiss} showed  that there is {\it exactly one} such measure (i.e., the flow on $\mathcal{M}$ is uniquely ergodic). 
Their proof is based on an ergodic argument.
Let us denote this measure by $\mu$. I  shall
refer to this measure  as in \cite{Fouchrandomizer} as the {\it Glasner-Weiss measure}.

We write $\mathcal{M}_f$ for the set of finite total orders on some subset of $\mathbb{N}$. For $\ell \in \mathcal{M}_f$,  denote by $Z_\ell$ the set of $\xi \in \mathcal{M}$, such that $\xi$ is an extension of $\ell$. These  sets are the cylinder subsets of $\mathcal{M}$. Write $\mathcal{Z}_0$ for the class of events of the form $Z_\ell$ for some $\ell \in \mathcal{M}_f$ and $\mathcal{Z}$ for the algebra generated by $\mathcal{Z}_0$. Note that the $\sigma$-algebra generated by
 $\mathcal{Z}$ is exactly the Borel algebra on $\mathcal{M}$.

For $Z \in \mathcal{Z}_0$ we write $Z^0$ for the complement of $Z$ and $Z^1$ for $Z$. Let  $(T_i)_{i \in \mathbb{N}}$ be any enumeration of  the algebra $\mathcal{Z}$ generated by $(Z_\ell)_{\ell \in \mathcal{M}_f}$ in such a way that one can effectively retrieve from  a given $i \in \mathbb{N}$,  the corresponding  $T_i$ as a finite union of sets $T$ of the form 
\begin{equation}
T = Z_{\ell_1}^ {\delta_1} \cap \ldots Z_{\ell_k} ^{\delta_k},
\label{eq:boolean}
\end {equation}
where each $\ell_i$ is in $\mathcal{M}_f$ and $\delta_i \in \{0,1\}$ for $i=1, \ldots, k$. We call any such enumeration a {\it recursive representation} of $\mathcal{Z}$.

The Glasner-Weiss measure $\mu$ is computable in the following sense: 

\begin{theorem} 
Denote by $\mu$ the Glasner-Weiss measure on the Borel-algebra of $\mathcal{M}$. Let $\left(T_{i} : i <\omega\right)$ be a recursive representation of the algebra $\mathcal{Z}$ . 
 There is an effective procedure that yields, for $i,k \in \mathbb{N}$,  a binary rational $\beta_{k}$ such that
$$|\mu\left(T_{i}\right)-\beta_{k}|<2^{-k}.$$
\end{theorem}

A proof of this result appears in \cite{Fouchrandomizer}.


\begin{definition}
 A set $A\subset \mathcal{M}$ is of \emph{constructive measure} $0$, if, for some recursive representation of  $\left(T_{i}:i \in \mathbb{N}\right)$ of $\mathcal{Z}$, there is a total recursive $\phi : \mathbb{N}^{2}\rightarrow \mathbb{N}$ such that
$$A\subset \bigcap_{n}\bigcup_{m}T_{\phi(n,m)}$$
and $\mu\left(\bigcup_{m}T_{\phi(n,m)}\right)$ converges effectively to $0$ as $n\rightarrow \infty$.
\end{definition}
\begin{definition}
A total order $\xi$ is said to be $\mu$-{\emph Martin-L\" of random}  if $\xi$ is in the complement of every subset $B$ of $\mathcal{M}$ of constructive measure $0$. 
\end{definition}

\section{The main theorem}

Write $ML_\mu \subset \mathcal{M}$ for the set  of  $\mu$-Martin-L\"of random total orders. 
\begin{theorem}
Let $\mathbf{K}$ be a recursive Fra\"\i ss\'e order class which is Ramsey and has the  ordering property. Write 
$$\mathbb{F} =(\mathbb{F}_0,<)$$ for its Fra\"\i ss\'e limit and $X_\mathbf{K}$ for the associated discerning flow.  Fix some recursive representation of $\mathbb{F}$. Note that
$$X_\mathbf{K} \subset \mathcal{M}.$$ If some  element of $X_\mathbf{K}$ is $\mu$-Martin-L\"of random, then the automorphism group $\mbox{Aut}(\mathbb{F}_0)$ is amenable. Equivalently, if  $\mbox{Aut}(\mathbb{F}_0)$ is not amenable, then
$$ML_\mu \cap X_\mathbf{K} = \emptyset.$$ 
Consequently, if for some $\xi \in X_\mathbf{K}$ and some automorphism $\pi$ of $\mathbb{F}_0$ it is the case that the linear order $\pi \xi$  is  $\mu$-Martin-L\"of random, then $\mbox{Aut}(\mathbb{F}_0)$ is  amenable.
\end{theorem}

Proof: Note that a topological group $G$ is amenable iff its universal minimal flow $M(G)$ has a $G$-invariant probability measure. Indeed, let $\nu$ be an invariant measure on $M(G)$. Consider any $G$-flow on some compact Hausdorff space $X$. By Zorn's lemma there is a minimal subflow $Y$ and a $G$-embedding $i$ of $Y$ into $X$. Therefore, there  are $G$-morphisms
\[\bfig
    \morphism (0,0) <500,0>[M(G)`Y;\pi]
     \morphism (500,0)/{>->}/<450,0>[Y`X; i]
\efig.\]
Let $\rho$ be the pushout measure of $\nu$ under $i\pi$.  In other words, for every Borel subset $A$ of $X$, we set
\[\rho(A) = \nu(\pi^{-1}i^{-1}A).\]
Then $\rho$ is an invariant measure on $X$. The converse is trivial, since $M(G)$ is a compact $G$-flow. 

We introduce a number of (standard) recursion-theoretic concepts and terminology: A sequence $\left(A_{n}\right)$ of sets in $\mathcal{Z}$ is said to be \emph {enumerable} if for each $n$, the set $A_n$  is of the form $T_{\phi(n)}$ for some total recursive function $\phi : \omega \rightarrow \omega$ and some effective enumeration $\left(T_{i}\right)$ of $\mathcal{Z}$.  (Note that the sequence $\left(A_{n}^{c}\right)$, where $A_{n}^{c}$ is the complement of $A_{n}$, is also an $\mathcal{Z}-$enumerable sequence.) In this case, we call the union $\bigcup_{n}A_{n} $ a $\sum_{1}^{0}$ set.  A set is a $\prod_{1}^{0}$ set if it is the complement of a $\sum_{1}^{0}$ set.  It is of the form $\bigcap_{n}A_{n}$, for some $\mathcal{Z}$-semirecursive sequence $\left(A_{n}\right)$.

We shall also need the following observation. (In the language of algorithmic randomness, it states the  well-known fact that the notion of  Martin-L\"of randomness is stronger than that of Kurtz randomness. A proof of this observation, in the present context, can be found in \cite{Fouchrandomizer}. For more on Kurtz randomness, the reader is referred to the book \cite{DowneyHirschfeldt} by Downey and Hirschfeldt.)
\begin{lemma}
If $A$ is a $\Sigma_{1}^{0}$ subset of $\mathcal{M}$  and if $\mu(A)=1$, then $ML_\mu$ is contained in $A$. In particular, if $B$ is a $\Pi_1^0$ subset of $\mathcal{M}$ that contains some element of $ML_\mu$, then $\mu(B) >0.$ 
\label{prop:kurtz}
\end{lemma}

It follows from Proposition \ref{lemmacrux} that a total order $\xi$ belongs to $X_\mathbf{K}$ iff for any $A$  in the age of $\mathbb{F}_0$ it is the case that $<^A$ is the restriction of $\xi$ to $A$. Therefore, since $\mathbf{K}$ is a {\it recursive} order class, the relation
$$\xi \in X_\mathbf{K}$$
is $\Pi_1^0$ definable over $\mathcal{M}$.  It follows from Lemma \ref{prop:kurtz} that, if $$ML_\mu \cap X_\mathbf{K} \neq \emptyset,$$ then $\mu(X_\mathbf{K})>0.$ This means that $\mu$ is a {\it nonzero} $\mbox{Aut}(\mathbb{F}_0)$-invariant measure on the flow $X_\mathbf{K}$.

Since $\mathbf{K}$ is a Ramsey order class, $X_\mathbf{K}$ is the {\it universal} minimal flow associated with the group $\mbox{Aut}(\mathbb{F}_0)$. By universality we can conclude that any $\mbox{Aut}(\mathbb{F}_0)$-flow on a compact Hausdorff space will admit a nonzero $\mbox{Aut}(\mathbb{F}_0)$-invariant measure. In particular,  $\mbox{Aut}(\mathbb{F}_0)$ is an amenable topological group.

The second part now follows from the observation that if $\xi \in X_\mathbf{K}$ and $\pi$ is an automorphism of $\mathbb{F}_0$ then $\pi \xi$ will also belong to $X_\mathbf{K}$.
\begin{corollary} 
Fix a recursive representation of the universal poset $\mathbb{P}_0$ on the natural numbers $\mathbb{N}$. Let $\mathcal{M}(\mathbb{P}_0)$ be the class of linear extensions of $\mathbb{P}_0$. Write $ML_\mu$ for the set of total orders on $\mathbb{N}$ that are Martin-L\" of random relative to the Glasner-Weiss probability measure $\mu$. Then
\[ML_\mu \cap \mathcal{M}(\mathbb{P}_0) = \emptyset.\]
\label{th:posetextensions}
\end{corollary}
Proof: The result is a direct consequence of the fact  that $\mbox{Aut}(\mathbb{P}_0)$ is not an amenable group. \cite{KechrisSokic}.

\section{Open problems}
 The following theorem is in \cite{Fouchrandomizer}.
\begin{theorem}
Write $\mathcal{Q}$ for the set of total orders on $\mathbb{N}$ which are isomorphic to the Cantor rational order $\eta$. Then
\[ML_\mu \subset \mathcal{Q}.\]
In particular,
\[\mu(\mathcal{Q})=1.\]
\label{th:mlcantor}
\end{theorem}
This observation has the following consequence. 

\begin{theorem}
For a total order $\eta$, set
\[S_\mu(\eta): =\{\sigma \in S_\infty: \sigma\eta \in ML_\mu\}.\]   Then $S_\mu(\eta) \neq \emptyset$ iff $\eta$ is a rational Cantor order.
\end{theorem}
Proof. By Theorem \ref{th:mlcantor},  if $\eta$ were not rational, the corresponding set $S_\mu(\eta)$ must be the empty set. 
If $\eta$ is rational, then the class $\mathcal{Q}$ is exactly the orbit of $\eta$ under the action of $S_\infty$. Since both $\mathcal{Q}$ and $ML_\mu$ have $\mu$-measure one, it follows that
$$\mu(\mathcal{Q} \cap ML_\mu)=1,$$
and, therefore, that $S_\mu(\eta) \neq \emptyset$.

Following \cite{Fouchrandomizer}, note that, if $\pi \in S_\infty$, then 
\begin{equation}
 S_\mu(\eta)\pi^{-1} = S_\mu(\pi\eta).
\label{eq:symmetrizers}
\end{equation}
 Indeed, for $\alpha \in S_\mu(\eta)$, we have $\alpha\pi^{-1}(\pi\eta) =\alpha\eta \in ML_\mu$ and hence $\alpha\pi^{-1} \in  S_\mu(\pi\eta)$.  Conversely, if $\tau \in S_\mu(\pi\eta)$, then $\tau\pi\eta \in ML_\mu$, i.e., $\tau \pi \in  S_\mu(\eta) $, and, so, $\tau \in S_\mu(\eta)\pi^{-1}$. 
 
 If $\eta_1,\eta_2 \in \mathcal{Q}$, there is some $\pi \in S_\infty$ such that $\eta_2=\pi\eta_1$. Moreover, if $\eta_1,\eta_2$ 
 were both recursive, the permutation $\pi$ could also be chosen to be recursive. (See Theorem \ref{th:efffraisse}). Write $S_r$ for the class of recursive permutations of $\mathbb{N}$. We let $S_r$ act on the right on the class $\Sigma$ of all sets of the form  $S_\mu(\tau)$ with $\tau$ a recursive rational order on $\mathbb{N}$.  The action is given by
 $$\Sigma \times S_r \longrightarrow \Sigma,$$
 $$(S_\mu(\tau), \pi) \mapsto S_\mu(\tau)\pi^{-1}, \; \pi \in S_r,\; \tau \in \mathcal{Q}_r,$$ where  $\mathcal{Q}_r$ denotes the class of all recursive rational orders on $\mathbb{N}$. It follows from the preceding arguments that this $S_r$-action will have a single orbit, i.e, the action is transitive. 
 Set
 $$\mathcal{S} =\bigcup_{\tau \in \mathcal{Q}_r} S_\mu(\tau).$$ If we choose any fixed $\eta \in \mathcal{Q}_r$, we also have
 $$\mathcal{S} =\bigcup_{\pi \in S_r} S_\mu(\eta)\pi^{-1}.$$
 We shall call the permutations in $\mathcal{S}$ {\it Martin-L\" of randomizers}. These are the permutations that transform some recursive rational order to one which is $\mu$-Martin-L\" of random.  

These arguments show that an understanding of $\mathcal{S}$ can be be attained from any single $ S_\mu(\tau)$  for a single recursive rational order $\tau$ modulo  the recursive permutations in $S_\infty$.

 Let  $\mathbf{K}$ be a recursive Fra\"\i ss\'e order class which is Ramsey and has the  ordering property. Write  again 
$$\mathbb{F} =(\mathbb{F}_0,<)$$ for its Fra\"\i ss\'e limit and $X_\mathbf{K}$ for the associated discerning flow.  The arguments of this paper show that, if $\tau \in \mathcal{Q}_r \cap X_\mathbf{K}$, then the presence of elements in $\mbox{Aut}(\mathbb{F}_0)$  which are Martin-L\" of randomizers of $\tau$ is related to the amenability of the group $\mbox{Aut}(\mathbb{F}_0)$. Indeed, for some $\pi \in \mbox{Aut}(\mathbb{F}_0)$ to be a Martin-L\" of randomizer of any $\tau$ as above is a {\it generic}  property, in the sense that this very fact  forces the group $\mbox{Aut}(\mathbb{F}_0)$ to be amenable! The problem still remains to identify the  class of Martin-L\" of randomizers.
\item Let $\mathbf{L}$ be the Fra\" {i}ss\' e  order class consisting of all pairs $(L,<)$ where $L$ is a lattice with underlying set  a finite ordinal and with $<$ being a total order on the underlying set of $L$ which is a linear extension of the partial order on $L$. As far as the author knows, it is unknown whether $\mathbf{L}$ is Ramsey and whether it has the ordering property. The author has discussed this problem with specialists in Ramsey theory and it would appear that this problem is wide open.  Writing  $\mathbb{L}=(\mathbb{L}_0,<)$ for the Fra\" {i}ss\" e limit of $\mathbf{L}$, it is also an interesting open problem to relate $X_\mathbf{L}$ to  $ML_\mu$ and thus perhaps gaining an understanding of the amenability or not of $\mbox{Aut}(\mathbb{L}_0)$. Note that if $\mathbf{L}$ were Ramsey with the ordering property, then $\mbox{Aut}(\mathbb{L})$ would be an extremely amenable group. This would mean  that its universal minimal flow is a singleton.



\begin{thebibliography}{10}

\bibitem{DowneyHirschfeldt}
R. G. Downey and D. R. Hirschfeldt.
\newblock {\em Algorithmic Randomness and Complexity, Theory and Applications of Computability}.
\newblock Springer, 2010.

\bibitem{Ellis}
R.~Ellis.
\newblock Universal minimal sets.
\newblock {\em Proc. Amer. Math. Soc.}, 11:272--281, 1949.

\bibitem{Fouchposet}
W.~L. Fouch{\'e}.
\newblock {S}ymmetry and the {R}amsey degree of posets.
\newblock {\em Discrete Math}, 167/168:309--315, 1997.

\bibitem{FouchPot}
W.~L. Fouch{\'e} and P.~H. Potgieter.
\newblock {K}olmogorov complexity and symmetrical relational structures.
\newblock {\em J. Symbolic Logic}, 63:1083--1094, 1998.

\bibitem{Fouchrandomizer}
W.~L. Fouch{\'e}.
\newblock Martin-L\" of randomness, invariant measures and  countable homogeneous structures.
\newblock {\em Theory of Computing Systems}, (to appear).
\newblock Available at arXiv:1205:0386v1.



\bibitem{GlasnerWeiss}
E.~Glasner and B.~Weiss.
\newblock {M}inimal actions of the group ${S}(\mathbb{Z})$ of permutations of
  the integers.
\newblock {\em Geometric And Functional Analysis}, 5:964--988, 2002.


\bibitem{Hodges}
W.~Hodges.
\newblock {\em {A} shorter model theory}.
\newblock Cambridge University Press, 1993.

\bibitem{Hrushovski}
E.~Hrushovski.
\newblock Extending partial isomorphisms of graphs.
\newblock {\em Combinatorica}, 12:411--416, 1992.

\bibitem{Kechris2011}
A.~S. Kechris.
\newblock {T}he dynamics of automorphism groups of homogeneous structures.
\newblock Lecture at LMS Northern Regional Meeting, July 2011.

\bibitem{Kechrisetal}
A.~S. Kechris, V.~G. Pestov, and S.~Todorcevic.
\newblock {F}ra\"{i}ss{\'e} limits, {R}amsey theory, and topological dynamics
  of automorphism groups.
\newblock {\em Geometric And Functional Analysis}, 15:106--189, 2005.

\bibitem{KechrisRosendal}
A.~S. Kechris and C.~Rosendal.
\newblock {T}urbulence, amalgamation and generic automorphisms of homogeneous
  structures.
\newblock {\em Proc. London Math. Soc.}, 94(3):302--350, 2007.

\bibitem{KechrisSokic}
A.~S. Kechris and M.~Soki\v c.
\newblock {D}ynamical properties of the automorphism groups of the random poset
  and random distributive lattice.
\newblock  Available at http://www.math.caltech.edu/people/kechris.html,
  2011.




\bibitem{PetrovVershik1}
F.~Petrov and A.~Vershik.
\newblock {U}ncountable graphs and invariant means on the set of universal
  countable graphs.
\newblock {\em Random Structures and Algorithms}, 126:389--405, 2010.



\end{thebibliography}

\end{document}